\documentclass[11pt,a4paper]{article}
\usepackage[utf8]{inputenc}
\usepackage[left=1in,right=1in,top=1in,bottom=1in]{geometry}
\usepackage{amsmath}
\usepackage{amsthm}
\usepackage{amsfonts}
\usepackage{amssymb}
\usepackage{mathtools}
\usepackage{algorithm}
\usepackage[noend]{algpseudocode}
\usepackage{float}
\usepackage{multicol}
\usepackage{graphicx}
\usepackage{verbatim}
\usepackage{ctable}
\usepackage{authblk}
\usepackage{makecell}
\usepackage{fancyhdr}
\usepackage{color}
\usepackage{hyperref}
\usepackage[section]{placeins}
\begin{document}
\title{A Discrete Approximation to Gibbs Free Energy of Chemical Reactions is Needed for Accurately Calculating Entropy Production in Mesoscopic Simulations }
\author[1]{Carlos Floyd}
\author[1,2,3,$\dag$]{Garegin A. Papoian}
\author[1,2,3,4,*]{Christopher Jarzynski}
\affil[1]{Biophysics Program, University of Maryland, College Park, MD 20742 USA}
\affil[2]{Department of Chemistry and Biochemistry, University of Maryland, College Park, MD 20742 USA}
\affil[3]{Institute for Physical Science and Technology, University of Maryland, College Park, MD 20742 USA}
\affil[4]{Department of Physics, University of Maryland, College Park, MD 20742 USA}
\affil[*]{email: cjarzyns@umd.edu}
\affil[$\dag$]{email: gpapoian@umd.edu}
\date{\today}

\begin{titlepage}
\maketitle
\abstract{In modeling the interior of cells by simulating a reaction-diffusion master equation over a grid of compartments, one employs the assumption that the copy numbers of various chemical species are small, discrete quantities.  We show that in this case, textbook expressions for the change in Gibbs free energy accompanying a chemical reaction or diffusion between adjacent compartments become inaccurate.  We derive exact expressions for these free energy changes under the assumption of discrete copy numbers and illustrate how these expressions reduce to the textbook expressions under a series of successive approximations leveraging the relative sizes of the stoichiometric coefficients and the copy numbers of the solutes and solvent.  Numerical results are presented to corroborate the claim that if the copy numbers are treated as discrete quantities, then only these more exact expressions lead to correct equilibrium behavior.  The newly derived expressions are critical for correctly tracking dissipation and entropy production in mesoscopic simulations based on the reaction-diffusion master equation formalism.

}
\end{titlepage}

\section{Introduction}

\begin{figure}[H]
	\centering
	\includegraphics[width= 12 cm]{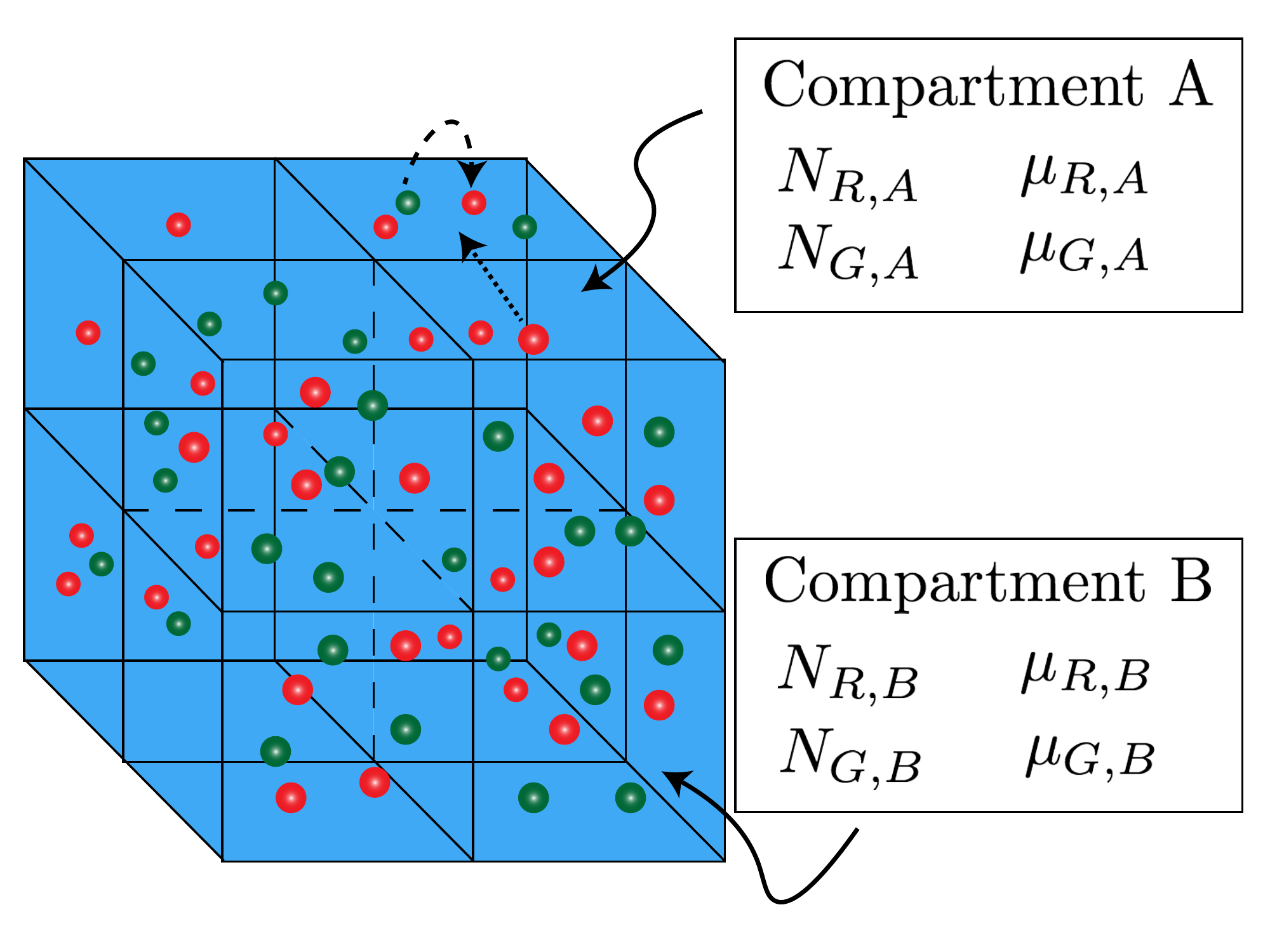}
	\caption{An example of a cubic compartment grid for use in simulation of a RDME.  Each compartment, labeled with letters $A, B, ...$, has local values of the quantities $N_R$, $N_G$, referring to the copy numbers of the red and green molecules in the compartment, and of $\mu_R$, $\mu_G$, referring to the chemical potentials of those molecules.  Molecules can react with each other within compartments (long dashed arrow), as well as hop between adjacent compartments, representing diffusion (short dashed arrow).}
	\label{cube}
\end{figure}

In recent years the coarse-grained computational modeling of intracellular environments has enjoyed significant advances.  An important paradigm shared by many such models is to treat the evolution of reacting chemical species' copy numbers and spatial distributions by simulating a reaction-diffusion master equation (RDME) \cite{grima2008modelling}.  In this approach, the system volume is divided into compartments, each with local values of the copy numbers and chemical potentials of the different chemical species (Figure \ref{cube}).  The RDME is a differential equation describing the evolution of the probability $P(\mathbf{N},t)$ of observing the vector of copy numbers $\mathbf{N} = \{N_{i,A}\}_{i \in S, A \in E}$ of chemical species $i$ in compartment $A$ at time $t$, where $S$ is the set of solute species and $E$ is the set of compartments in the system.  The RDME can be written schematically as 
\begin{equation}
	\frac{dP\left(\mathbf{N}, t \right)}{dt} = \left(\hat{M} + \hat{D}\right)P\left(\mathbf{N},t\right)
	\label{rdme}
\end{equation}
where $\hat{M}$ and $\hat{D}$ represent operators describing chemical reactions and inter-compartment diffusion, respectively \cite{tanaka2010reverse}.  An in-depth description of the RDME approach can be found in \cite{baras1996reaction, bernstein2005simulating}, where the form of the operators is discussed.  As opposed to directly solving Equation \ref{rdme}, one often simulates trajectories of the vector $\mathbf{N}$ obeying the stochastic dynamics encoded in the RDME, using a variant of the Gillespie algorithm \cite{gillespie1977exact}.  The sizes of the compartments are commonly determined by the Kuramoto lengths, the mean free diffusional path for a species before it participates in a chemical reaction.  Within each compartment, the spatial distribution of the reacting species is assumed to be homogeneous, allowing the use of mass-action kinetics using the compartment's local values of the species' concentrations  to describe the stochastic chemical reaction propensities.  Molecules can additionally jump between adjacent compartments in ``diffusion events" (whose propensities also depend on the compartments' local concentrations of species) to give rise to concentration gradients on the scale of the compartment length.  This modeling approach is appropriate when the Kuramoto length $l_K$ is small compared to the system size (i.e. the assumption of homogeneity over the system volume fails), but is much larger than the intermolecular distance scale \cite{grima2008modelling}.  
Examples of simulation platforms employing such an approach include Virtual Cell \cite{loew2001virtual, slepchenko2003quantitative}, lattice microbes \cite{roberts2013lattice}, and MEDYAN (Mechanochemical Dynamics of Active Networks) \cite{popov2016medyan}.
  
One important aspect of simulating non-equilibrium biological systems is the computation of thermodynamic forces that drive the observed flux on the network of chemical reactions \cite{hill2004free, earnest2018simulating, beard2007relationship}.  Determining these forces can allow for the quantification of entropy production in chemically reactive systems \cite{prigogine1967introduction}.  In several research groups, measuring entropy production in biological active matter has been a recent goal \cite{england2015dissipative, seara2018entropy}.  For instance, in an accompanying article under preparation by the authors of this paper, we quantify the entropy production rates of self-organizing non-equilibrium actomyosin networks in MEDYAN using the expressions derived here as a first step.  The ability to measure dissipation in active matter systems will allow to test the applicability of different physical organizing principles relating the production of entropy to the likelihood of observing certain trajectories \cite{england2013statistical,prigogine1971biological}.  For isothermal, isobaric, chemically reactive solutions, which includes many biological systems, measuring the total entropy production amounts to determining the change in Gibbs free energy that accompanies chemical reactions and diffusion down concentration gradients \cite{callen1998thermodynamics, attard2002thermodynamics, nicolis1977self}.  A ubiquitous textbook expression for the change in Gibbs free energy $G$ accompanying a chemical reaction is 
\begin{equation}
\Delta G = k_B T \log K_{eq} + k_B T \log Q
\label{eqfirst}
\end{equation}
where $K_{eq}$ is the equilibrium constant, $Q$ is the reaction quotient, $k_B$ is Boltzmann's constant, and $T$ is the temperature.  At equilibrium, $Q = K_{eq}^{-1}$ and as a result $\Delta G = 0$.  However, in this paper, we argue that Equation \ref{eqfirst} is a biased approximation to the exact value of $\Delta G$ accompanying a chemical reaction which holds when the copy numbers of the reacting molecules are on the order of Avogadro's number.  When  the system is small, such as when copy numbers are on the order of 100 as is the case for certain reacting molecules in intracellular reaction networks, then thermodynamic expressions such as Equation \ref{eqfirst} require corrections \cite{hill2001perspective, hill1994thermodynamics}.  As a simple motivating example, consider a mixture of an even number of two chemical species, red and green, which inter-convert at equal rates.  At equilibrium, the copy numbers of these molecules will be equal, and thus $Q = K_{eq}^{-1} = 1$.  Now if a reaction were to occur at equilibrium to produce an additional red molecule and one fewer green molecule, then we would expect that the Gibbs free energy of the system had increased, since we have left equilibrium where the free energy attains its minimum.  However the prediction of Equation \ref{eqfirst} is that $\Delta G = 0$ for this reaction.  The assumption whose violation leads to Equation \ref{eqfirst} becoming incorrect is that the copy number of chemical species is a continuous quantity.  When these variables are considered as discrete, then a different expression for $\Delta G$ must be used to give correct behavior.
 
Similarly for diffusion between adjacent compartments, a common expression for the change in Gibbs free energy accompanying the jump of a molecule $i$ from compartment $A$ with copy number $N_{i,A}$, to compartment $B$ where its copy number  is $N_{i,B}$, is 
\begin{equation}
\Delta G = k_B T \log \frac{N_{i,B}}{N_{i,A}}.
\label{eqsecond}
\end{equation}
Imagine however we have a situation where $N_{i,A} = N_{i,B}$, and a molecule jumps from compartment $A$ to $B$.  The Gibbs free energy should have increased since we have departed from the highest entropy distribution of the molecules over the two compartments, however Equation \ref{eqsecond} will predict that $\Delta G = 0$.    

In this paper we derive nearly exact expressions for the change in Gibbs free energy accompanying chemical reactions within compartments and diffusion events between compartments, and we further show how these expressions relate to the familiar textbook formulas Equations \ref{eqfirst} and \ref{eqsecond} through a series of approximations.  We also discuss the assumptions that go into defining the Gibbs free energy of a grid of homogeneous compartments which can exchange energy and particles, such as in a simulation of the RDME.  Finally we present numerical simulations using MEDYAN to demonstrate the need to use these more exact expressions for $\Delta G$ in order to obtain sensible results when copy numbers are treated as discrete variables.  Only these more exact expressions will give correct, unbiased behavior when measuring entropy production in mesoscopic \textit{in silico} studies of biological non-equilibrium systems which rely on the RDME formalism.

\section{Methods}
\subsection{$\Delta G$ of Reactions}
Here we make successive approximations to the formula for $\Delta G$ accompanying a chemical reaction, and our notation reflects the level of approximation in which certain variables are being used: where appropriate, we subscript variables with a parenthesized number, i.e. $\Delta G_{(0)}$, where increasing numbers represent more approximate versions.  The symbol $\widetilde{ \ \ \ \  }$ will indicate that the quantities of chemical species are being represented by mole fractions $\chi_i$, as opposed to by concentrations $C_i$.  In this section we treat the case that our system comprises a single closed  compartment of a homogeneous dilute solution in which a chemical reaction has occurred and derive an expression for the change in the Gibbs free energy.  In this system the number of solvent molecules is fixed and the solute molecules participate in chemical reactions, causing their copy numbers to change.  In the next sections we consider a system with multiple weakly interacting compartments, each of which represents a homogeneous solution with local copy numbers of solvent and solutes and between which both solvent and solutes can diffuse.  The nearly exact result Equation \ref{eqa16} obtained in this section will also apply to those systems, as we argue below.

Before restricting to the case of a single closed compartment, we establish notation for properties of the chemical species in a compartment grid.  The chemical potential for species $i$ in compartment $A$ can be expressed either as depending upon the mole fraction, $\chi_{i, A}$, or upon the concentration, $C_{i,A}$, of that species in the compartment: 
\begin{equation}
\label{eqa1}
\mu_{i, A} = \widetilde{\mu_i^0}(T,p) + k_B T \log{\chi_{i, A}} = \mu_i^0(T,p) + k_B T \log{C_{i, A}} 
\end{equation}  
where $k_B$ is Boltzmann's constant, $\widetilde{\mu_i^0}(T,p)$ is the standard state chemical potential at temperature $T$ and pressure $p$ when working with the unitless $\chi_{i, A}$, and $\mu_i^0(T,p)$ is the same when working with $C_{i,A}$.  $C_{i,A}$ and $\chi_{i,A}$ both play the role of the ``composition variable" leading to these different, yet equivalent expressions for the chemical potential \cite{de1986free}.  We make the distinction between dependence upon copy number and dependence on concentration in order to establish parameters that can be used in simulation, which commonly works with copy numbers, based on those given in the literature, which typically use units of concentration.  Here we make the assumption of an ideal dilute solution (i.e the solvent obeys Raoult's law and the solute obeys Henry's law), and we further neglect the coefficient of activity of the different species \cite{engel2006physical}.  The mole fraction can be written 
\begin{equation}
\label{eqa2}
\chi_{i, A} = \frac{N_{i,A}}{N_A} = \frac{N_{i,A}}{\sum_{j \in M} N_{j,A}}
\end{equation}
where $N_{i,A}$ is the copy number of species $i$ in compartment $A$, ${N_A}$ is the total copy number of molecules in compartment $A$, and $M$ is the set of all species including the solvent in the system.  Similarly, the concentration can be written 
\begin{equation}
\label{eqa3}
C_{i,A} = \frac{N_{i,A}}{\Theta_A}
\end{equation}
where
\begin{equation}
\label{eqa4}
\Theta_A = N_\text{Av} V_A
\end{equation}
is a conversion factor, $N_\text{Av}$ is Avogadro's number, and $V_A$ is the compartment volume.  Using Equations \ref{eqa1}, \ref{eqa2}, and \ref{eqa3}, we can write 
\begin{equation}
\label{eqa5}
\widetilde{\mu_i^0}(T,p) = \mu_i^0(T,p) + k_B T \log{\frac{N_A}{\Theta_A}}.
\end{equation}

Through standard arguments involving extensivity, one can establish that the Gibbs free energy of the solution in compartment $A$ can be written as a weighted sum over the chemical potentials of the species:
\begin{equation}
\label{eqa6}
G_A (\bold{N}_A) = \sum_{i \in M} N_{i,A}  \mu_{i,A} (\chi_{i,A}) 
\end{equation} 
where $\bold{N}_A = \{ N_{i,A}\}_{i \in M}$ is the vector of species copy numbers $N_{i,A}$, and where we have explicitly indicated the dependency of $\mu_{i,A}$ upon mole fraction $\chi_{i,A}$ via Equation \ref{eqa2} \cite{attard2002thermodynamics, anderson2014foundations}.  We rely on Equations \ref{eqa1} and \ref{eqa6} to derive changes in Gibbs free energy accompanying chemical reactions and inter-compartment diffusion.  

Consider a reaction of the general form
\begin{equation}
\label{eqa7}
\nu_1 X_1 + \nu_2 X_2 + \ldots \rightleftharpoons \upsilon_1 Y_1 + \upsilon_2 Y_2 + \ldots 
\end{equation}
where $X_i$ represent reactants, $Y_j$ represent products, $\nu_i$ and $\upsilon_j$ are stoichiometric coefficients, and the rate of reaction is $k_+$ to the right and $k_-$ to the left.  We have dropped the subscript $A$ indicating the compartment in which the reaction takes place, and now assume that our system is a single closed compartment.  When this reaction has occurred once to the right, the copy numbers of reactants have changed $N_i \rightarrow N_i - \nu_i$, and those of the products have changed $N_j \rightarrow N_j + \upsilon_j$.  We calculate the change in Gibbs free energy accompanying this reaction by considering it as resulting from these discrete changes in copy numbers \cite{hill1994thermodynamics}.  Using Equations \ref{eqa1}, \ref{eqa2}, and \ref{eqa6}, the Gibbs free energy before the reaction has occurred can be written as 

\begin{align}
G^\text{initial} &= \sum_{i \in R} N_i\left(\widetilde{\mu_i^0} + k_B T \log{\frac{N_i}{N}}\right) + \sum_{j \in P} N_j\left(\widetilde{\mu_j^0} + k_B T \log{\frac{N_j}{N}}\right) \nonumber \\
&+ N_s \left(\widetilde{\mu_s^{0 *}} + k_B T \log{\frac{N_s}{N}}\right) \label{eqa8}
\end{align}
where $R$ is the set of reactants, $P$ is the set of products, the subscript $s$ refers to the solvent, and we have dropped the dependence of the standard state chemical potential on $T$ and $p$.  Note that $\widetilde{\mu_i^{0}}$,  $\widetilde{\mu_j^{0}}$ describe the chemical potential at a reference concentration of the solute in the solvent (also referred to as the solute standard state), whereas $\widetilde{\mu_s^{0*}}$ describes the chemical potential at a reference state of pure solvent (also referred to as the solvent standard state) \cite{engel2006physical}.  We assume here for simplicity and without loss of generality that there are no solute species which have not participated in the reaction (i.e. spectator solute species).  These species would also contribute terms to Equation \ref{eqa8} but, when we subtract the initial from the final Gibbs free energy, their inclusion would not give a different result.  The final Gibbs free energy is 
\begin{align}
G^\text{final} &= \sum_{i \in R} (N_i-\nu_i)\left(\widetilde{\mu_i^0} + k_B T \log{\frac{N_i-\nu_i}{N+\sigma}}\right) + \sum_{j \in P} (N_j+\upsilon_j)\left(\widetilde{\mu_j^0} + k_B T \log{\frac{N_j+\upsilon_j}{N+\sigma}}\right) \nonumber \\
&+ N_s \left(\widetilde{\mu_s^{0*}} + k_B T \log{\frac{N_s}{N+\sigma}}\right) \label{eqa9}
\end{align}
where 
\begin{equation}
\sigma = \sum_{j \in P} \upsilon_j - \sum_{i \in R} \nu_i \label{eqa10}
\end{equation}
is what we refer to as the ``stoichiometric difference", or the amount by which the total species copy number $N$ has changed.  As described in \cite{de1986free,barrow1983free}, when $\sigma \neq 0$ it is important to account for the solvent species in the calculation of free energy differences.  This is because the free energy of the solvent, which is the most numerous species in the reaction volume, will not be the same after the reaction has taken place since its mole fraction will change as $N$ changes to $N + \sigma$.  Neglecting the solvent species when $\sigma \neq 0$ leads to expressions for $\Delta G$ that are off by an amount $\sigma k_B T$ \cite{de1986free}.  Note that, whereas those authors describe the appearance of this erroneous term while formulating the Gibbs free energy as a function of a continuous degree of advancement of reaction $d\xi = -\frac{d N_i}{ \nu_i} = \frac{d N_j}{ \upsilon_j}$, here we treat the extent of reaction as a discrete quantity.  In the limit that $\frac{\nu_i}{N_i} \rightarrow 0$, $\frac{\upsilon_j}{N_j} \rightarrow 0$ for each reactant and product, the discrete case passes into the continuum case, however under the assumption of small copy numbers we do not satisfy this limit.  In Appendix A we discuss further differences between the continuum treatment and the discrete treatment, as well as the relation to the Gibbs-Duhem Equation.  

After some algebra (using the fact that $\sum_{i \in R} N_i + \sum_{j \in P} N_j + N_s = N$), we can write the change in Gibbs free energy as 
\begin{align}
\Delta G_{(0)} &= G^\text{final} - G^\text{initial} \nonumber \\
&= \widetilde{\Delta G^0} + k_B T \log  \prod_{i \in R} \frac{(N_i - \nu_i)^{N_i-\nu_i}}{N_i^{N_i}}  \prod_{j \in P} \frac{(N_j + \upsilon_j)^{N_j+\upsilon_j}}{N_j^{N_j}} \frac{N^N}{(N+\sigma)^{N+\sigma}}  \label{eqa11}
\end{align} 
where 
\begin{equation}
\label{eqa12}
\widetilde{\Delta G^0} = \sum_{j \in P} \upsilon_j \widetilde{\mu_j^0} - \sum_{j \in R} \nu_i \widetilde{\mu_i^0}.  
\end{equation}
Equation \ref{eqa11} is exact, but we would like to avoid specifying $N$ in simulation since the solvent is typically not modeled explicitly, as we elaborate on in the next section.  We would also like a way to find $\widetilde{\Delta G^0}$ based on literature values of $\Delta G^0$.  To these ends we first rewrite Equation \ref{eqa11} as 
\begin{equation}
\label{eqa13}
\Delta G_{(0)} = \widetilde{\Delta G^0} + k_B T \log{\widetilde{Q}_{(0)}} + k_B T \log{\frac{N^N}{(N+\sigma)^{N+\sigma}}}
\end{equation}
where 
\begin{equation}
\label{eqa14}
\widetilde{Q}_{(0)} = \prod_{i \in R} \frac{(N_i - \nu_i)^{N_i-\nu_i}}{N_i^{N_i}}  \prod_{j \in P} \frac{(N_j + \upsilon_j)^{N_j+\upsilon_j}}{N_j^{N_j}}.
\end{equation}
From Equations \ref{eqa5} and \ref{eqa12} we can write 
\begin{equation}
\widetilde{\Delta G^{0}} = \Delta G^0 + \sigma k_B T \log{\frac{N}{\Theta}}
\end{equation}
where 
\begin{equation}
\Delta G^0 = \sum_{j \in P} \upsilon_j \mu_j^0 - \sum_{j \in R} \nu_i \mu_i^0. 
\end{equation}
Inserting this to Equation \ref{eqa13}, we have 
\begin{align}
\Delta G_{(0)} &= \Delta G^0 + k_B T \log{\left(\frac{N}{\Theta}\right)^\sigma \frac{N^N}{(N+\sigma)^{N+\sigma}}} + k_B T \log{\widetilde{Q}_{(0)}} \nonumber\\
&= \Delta G^0 - \sigma k_B T \log{\Theta} + k_B T \log{ \bigg(\frac{N}{N+\sigma}\bigg)^{N+\sigma}} + k_B T \log{\widetilde{Q}_{(0)}}. \label{eqa15}
\end{align}
We now make the approximation that $N \gg \sigma$, which is certainly reasonable (in the example of a $0.125$ $\mu m^3$ compartment filled with water, $N$ is on the order of $10^9$ and $\sigma$ is on the order of $1$), to write the third term in Equation \ref{eqa15} as $-\sigma k_B T$ \footnote{We have $-\log \left(\frac{N+\sigma}{N}\right)^{(N+\sigma)} = -\log \left(1 + \frac{\sigma}{N}\right)^{(N+\sigma)} \approx -\sigma$ for large $N$.}, giving
\begin{equation}
\label{eqa16}
\Delta G_{(1)} = \Delta G^0 - \sigma k_B T  \log{\Theta} - \sigma k_B T + k_B T \log{\widetilde{Q}_{(1)}},
\end{equation}
where $\widetilde{Q}_{(1)} = \widetilde{Q}_{(0)}$ (we update the subscript to indicate the use of this quantity in a more approximate version of the formula for $\Delta G$).  We recommend using Equation \ref{eqa16} in simulation because it allows to incorporate literature values for $\Delta G^0$ and avoids specification of $N$.  The combination $\Delta G^0 - \sigma k_BT \log{\Theta}$ has units of energy, since if $\sigma$ is non-zero, then the equilibrium constant $K_{eq}$ will not be unitless, causing $\Delta G^0 = k_B T \log K_{eq}$ to not be unitless unless the factor $\sigma k_BT \log{\Theta}$ is subtracted from it.  To understand the term $-\sigma k_B T$ in Equation \ref{eqa16}, we make further approximations leveraging the large sizes of the solute copy numbers compared to their stoichiometric coefficients.  First, we can rewrite $\widetilde{Q}_{(1)}$ as 
\begin{equation}
\label{eqa17}
\widetilde{Q}_{(1)} = \prod_{i \in R} \left(1-\frac{\nu_i}{N_i}\right)^{N_i}(N_i - \nu_i)^{-\nu_i}  \prod_{j \in P} \left(1+\frac{\upsilon_j}{N_j}\right)^{N_i}(N_j + \upsilon_j)^{\upsilon_j}.
\end{equation}
Assuming that $N_i \gg \nu_i$ and $N_j \gg \upsilon_j$, and using the limit
\begin{equation}
\label{eqa18}
\lim_{x\rightarrow \infty} \left(1\pm \frac{y}{x}\right)^x =  e^{\pm y}
\end{equation}
we approximate $k_B T \log{\widetilde{Q}_{(1)}}$ as 
\begin{equation}
k_B T \log{\widetilde{Q}_{(1)}} \approx \sigma k_B T + k_B T \log{  \prod_{i \in R} (N_i - \nu_i)^{-\nu_i}  \prod_{j \in P} (N_j + \upsilon_j)^{\upsilon_j}}.
\end{equation}
Inserting this into Equation \ref{eqa16}, canceling the factor of $\sigma k_B T$, gives
\begin{equation}
\label{eqa19}
\Delta G_{(2)} = \Delta G^0 - \sigma k_B T \log{\Theta} + k_B T \log{\widetilde{Q}_{(2)}},
\end{equation}
where 
\begin{equation}
\label{eqa20}
\widetilde{Q}_{(2)} = \prod_{i \in R}(N_i - \nu_i)^{-\nu_i}  \prod_{j \in P} (N_j + \upsilon_j)^{\upsilon_j}.
\end{equation}

A common textbook expression for the change in Gibbs free energy is 
\begin{equation}
\label{eqa21}
\Delta G_{(3)} = \Delta G^0 + k_B T \log{Q} = \Delta G^0 - \sigma k_B T \log \Theta + k_B T \log{\widetilde{Q}_{(3)}} 
\end{equation}
where 
\begin{equation}
\label{eqa22}
Q = \prod_{i \in R} C_i^{-\nu_i} \prod_{j \in P} C_j^{\upsilon_j}
\end{equation}
and 
\begin{equation}
\label{eqa23}
\widetilde{Q}_{(3)} = \prod_{i \in R} N_i^{-\nu_i} \prod_{j \in P} N_j^{\upsilon_j}.
\end{equation}
We see that Equation \ref{eqa21} is obtained from Equation \ref{eqa19} upon making the approximations $N_i - \nu_i \approx N_i$ and $N_j + \upsilon_j \approx N_j$.

We thus have four expressions for $\Delta G$ of chemical reactions:
\begin{itemize}
	\item Equation \ref{eqa11} is exact, however it requires specifying $N$.
	\item Equation \ref{eqa16} uses the approximation $N \gg \sigma$ and avoids specifying $N$.  We recommend the use of this expression because it is the most exact expression for which we need not specify $N$, and since it is written in terms of $\Delta G^0$, for which literature values can be found. 
	\item Equation \ref{eqa19} uses the approximations $N_i \gg \nu_i$ and $N_j \gg \upsilon_j$.
	\item Equation \ref{eqa21} uses the approximations $N_i \gg \nu_i$ and $N_j \gg \upsilon_j$ again.
\end{itemize}
In Appendix B, we provide expressions for the accuracy of these approximations.  Note that, without specify the copy number of solvents, we can only approximately compute changes in Gibbs free energy, and not the instantaneous Gibbs free energy of the system.  Typically only the changes are of interest.  When employing any of the above expressions which involve the logarithms of products of copy numbers that are raised to the power of other copy numbers, we recommend splitting the logarithm of products into a sum of logarithms in order to prevent overflow resulting from computing very large numbers.  These results are independent of the assumption of a system consisting of multiple, weakly-interacting compartments, which we discuss next.

\subsection{Thermodynamics of a Reaction-Diffusion Compartment Grid}
In simulating a RDME, one commonly treats diffusion between adjacent compartments and chemical reactions within compartments using an augmented set of all species and reactions in the system that treats species as distinct if they belong to separate compartments.  Thus if there are $|S|$ reacting species and $|E|$ compartments, where $S$ and $E$ are the sets of solute species and compartments respectively, then in the augmented set there are $|S||E|$ species tracked.  The solvent species is not tracked explicitly, as we elaborate on below.  The number of reactions in the augmented system, including $r$ chemical reactions per compartment and roughly $z|S|$ diffusion events per compartments (where $z$ is the assumed constant number of neighbors of each compartment, ignoring boundary compartments), is $|E|(r + z|S|)$.  This augmented set of species and reactions is then simulated using a Gillespie algorithm, where the propensities are appropriately scaled according to the volume of the compartments \cite{bernstein2005simulating}.

Crucial to the justification of this strategy to simulate a RDME is the assumption that within each compartment, the reacting species are considered homogeneously distributed so that one may use mass-action kinetics to determine the propensities.  This assumption amounts to the condition that the timescale describing diffusion within compartments, $\tau_D$, is much less than the timescale of chemical reactions, $\tau_C$:
\begin{equation}
\tau_D \ll \tau_C.
\label{eqn1}
\end{equation}
This comparison should be done for each diffusing and reacting species, and the timescale of the fastest reaction (taken as the inverse of the propensity) for each species should be used.  If the condition holds, then the process of intra-compartmental diffusion will homogenize the solution faster than chemical reactions occur, so the assumption of mass-action kinetics holds.  Let the dimension of the space be $d$, the length of the (here assumed cubical) compartments be $h$, and the diffusion constant of a species be $D$.  Then 
\begin{equation}
\tau_D \approx \frac{h^2}{2d D}.
\label{eqn2}
\end{equation}
The Kuramoto length is given by 
\begin{equation}
l_K = \sqrt{2 d D \tau_C},
\label{eqn3}
\end{equation} 
so one can see that the condition $\tau_D \ll \tau_C$ is equivalent to the condition $\l_K \gg h$, and thus one can enforce this condition by choosing a smaller compartment size $h$.  For fixed total volume, there is a trade-off between $h$ and $|E|$, which determines the size of the augmented system, and therefore the computational efficiency.  Note that the timescale of intra-compartment diffusion $\tau_D$ is approximately equal to the the timescale of inter-compartment diffusion (which can be given as the inverse of $k_D = \frac{D}{h^2}$), so a third way of describing this condition is that the frequency of jumps between adjacent compartments is much greater than the frequency of chemical reactions inside the compartments, which can be checked empirically in simulation \cite{grima2008modelling,bernstein2005simulating}.  We note that in the literature these expressions may differ up to a constant coefficient depending on the reference.      

In order to approximately describe the thermodynamics of this system, we distinguish between two types of components: the inert solvent and the dilute, chemically reactive solutes.  We assume here that the system is impermeable to the flow of either kind of species to the exterior.  Within the system, all species are permeable (though local diffusion constants may be incorporated for each species \cite{grima2008modelling}).  We would like to treat the solutes explicitly (at the level of their compartment concentrations), whereas we would like to model the solvent implicitly through an appropriate limit.  For instance, this was done to arrive at Equation \ref{eqa16} above.  We acknowledge that it is necessary to avoid tracking the changes in the solvent copy numbers in each compartment given the computational demand of doing so.  To this end we assume the existence of a laboratory timescale $\tau_l$ which is much longer than the timescale describing the local compartment fluctuations of the solvent $\tau_s$, yet shorter than but on the order of the timescale describing the diffusion of the solutes $\tau_D$.  On this timescale, then in the time between the chemical reactions and diffusion events involving the solutes, the system is quasi-equilibrated with respect to fluctuations involving the fast processes of solute and solvent homogenization, and thus the system may be assigned well-defined values of Gibbs free energy \cite{mishin2015thermodynamic}.  This hierarchy of timescales can be written as
\begin{equation}
\tau_s \ll \tau_l \lesssim \tau_D \ll \tau_C.
\label{eqn4}
\end{equation}
Typical ratios of the diffusion constants for solute to solvent are in the range of $1/10$ to $1/100$ \cite{milo2015cell}, placing $\tau_D / \tau_s$ in the range of $10$ to $100$.  The frequency of inter-compartment solvent diffusion will of course overwhelm the frequency of inter-compartment solute diffusion due to the comparative copy numbers of each.

Thus with the timescale $\tau_l$ there is enough resolution to track the diffusion and chemical reactions of the solutes, while allowing averaging over the fluctuations of the solvent.  To describe activity occurring over the entire grid of compartments for large systems, we introduce new timescales $\tau^g$.  If we hold $h$ and the concentrations fixed and add more compartments to our system, the rates of diffusion and chemical reactions occurring anywhere in the system will scale as the number of compartments $|E|$, and thus the timescales needed to describe them scale as $|E|^{-1}$.  The timescale of solvent fluctuations across the whole grid will also scale inversely with $|E|$, so Equation \ref{eqn4} does not ultimately depend on the number of compartments.  In other words, for systems with many compartments, then as long as Equation \ref{eqn4} holds for one compartment, we can be sure that our laboratory timescale that describes the whole grid $\tau_l^g \sim \tau_l / |E|$ will be short enough to describe activity of the solutes while long enough to allow averaging over the fluctuations of the solvent occurring locally in each compartment.

We assume that the exterior of the system acts as a reservoir for the thermodynamic variables $p$ and $T$.  The volume $V$ of the system also remains constant, however under the assumption of the solvent being an incompressible liquid, the change in quantity $pV$ is approximately zero, and for each reaction the change in Gibbs free energy is equal to that of the Helmholtz free energy.  Thus it is inconsequential whether we consider $p$ or $V$ to be reservoir variables, and we choose $p$ in order to speak of the Gibbs free energy of the system.  We can write the Gibbs free energy of the system as $G(\mathbf{N}, p, T)$, where $\mathbf{N} = \{ \{N_{i,A} \}_{i \in M} \}_{A \in E}$ represents the set of copy numbers of all solute and solvent species in each compartment in the grid.  We further assume the compartments to be only weakly-interacting; that is, they can exchange energy and particles, but the interaction of the two subsystems does not contribute a term to the Gibbs free energy of the system.  This is equivalent to assuming that the Gibbs free energy of the compartments $G_A$ are linearly additive: 
\begin{equation}
G = \sum_{A \in E} G_A
\end{equation}  
without any terms of the form $G_{AB}$.  To justify this, we first note that the interaction free energy between two adjacent compartments will primarily be due to the interaction of the solvent at the interface.  This interfacial free energy will, even for mesoscopically sized compartments, be small compared to the free energy due to the bulk of the compartment.  Further, our main interest will be in changes in the Gibbs free energy of the system due to solute diffusion between compartments and chemical reactions, neither of which will significantly affect the interfacial free energy, so all of the terms $G_{AB}$ will drop out of the expression for $\Delta G$.  

In our approach of averaging over fluctuations in the solvent amounts and taking the limit that this average is large compared to the copy numbers of the solutes, we are choosing to neglect the changes in Gibbs free energy of the system owing to the solvent fluctuations.  We justify this with an argument that these fluctuations are small compared to those resulting from the activity of the solute molecules, and also because it arises out of necessity due to the computational intractability of tracking the solvent fluctuations.  On the laboratory time scale $\tau_l$, each compartment has an average copy number of solvent molecules, $\overline{N_{s,A}}$.  The fluctuations in this quantity will have a standard deviation on the order of $\overline{N_{s,A}}^{1/2}$ \cite{mishin2015thermodynamic}.  As indicated previously (to arrive at Equation \ref{eqa16}), we avoid specifying $\overline{N_{s,A}}$ by taking the limit that it is much larger than the number of solute species, and thus fluctuations in this quantity don't matter when computing nearly exact changes in the Gibbs free energy accompanying reactions involving the solutes.  We need to establish that the change in Gibbs free energy of a compartment due to a fluctuation in the solvent copy number on the order of $\overline{N_{s,A}}^{1/2}$ is small compared to the change in Gibbs free energy accompanying chemical reactions and inter-compartment solute diffusion, and thus that the fluctuations in this quantity are not outweighing the changes that we are measuring.  In Appendix C we show that if the concentrations of solutes are very different in the two compartments, then this Gibbs free energy change is on the order of $\epsilon \overline{N_{s,A}}^{1/2} k_BT$, where $\epsilon$ is the ratio of solutes to solvent.  We show that this quantity is typically much less in magnitude than the change in Gibbs free energy accompanying a solute diffusion event.  If the concentrations are nearly equal, then the Gibbs free energy is on the order of $\epsilon k_B T$, and is thus negligibly small.  We make the modeling choice to ignore these changes in the Gibbs free energy resulting from solvent fluctuations because they tend to be small and they do not represent the processes we are interested in which involve the activity of the solute molecules. 

To summarize, we track the changes in the Gibbs free energy of the system by computing a value of $\Delta G$ whenever a chemical reaction occurs within a compartment or a solute diffusion event occurs between adjacent compartments.  By the linear additivity of the compartments' free energies, any change in the free energy of a single compartment is equal to the change in free energy of the whole system (i.e. $\Delta G_A = \Delta G$).  We assume that the activity of the solvent only contributes small, fluctuating, unbiased changes to the free energy which we ignore; we also assume that the amount of solvent is so large that one can neglect the fluctuations in this quantity when computing the changes in free energy for processes involving the solute copy numbers.  

\subsection{$\Delta G$ of Diffusion}
To describe the change in Gibbs free energy accompanying a solute diffusion event between neighboring compartments in a compartment-based reaction-diffusion scheme we use a similar approach to the one used above for chemical reactions.  The key difference here is that as opposed to multiple species being involved in a reaction taking place in a single compartment, we now have a single species involved in a reaction taking place between two compartments.  

Consider species $i$ diffusing from compartment $A$, where initially its copy number is $N_{i,A}$, to compartment $B$, where its copy number is $N_{i,B}$.  The total number of molecules in compartment $A$ is $N_A$, and in compartment $B$ it is $N_B$.  Assume there is just one spectator species constituting the solvent, labeled $s$ with copy numbers $N_{s,A}$ and $N_{s,B}$ (we incorporate our uncertainty in the exact values of these numbers by taking a limit later).  As a result of the diffusion event, we have the following changes in these quantities:
$N_{i,A} \rightarrow  N_{i,A} -1$, $N_{i,B} \rightarrow  N_{i,B} + 1$,$N_{s,A} \rightarrow N_{s,A}$, $N_{s,B} \rightarrow N_{s,B}$, $N_A \rightarrow N_A - 1$, and $N_B \rightarrow N_B + 1$.  The initial Gibbs free energy is 
\begin{align}
G^\text{initial} &= N_{i,A}\left(\widetilde{\mu_i^0} + k_B T \log  \frac{N_{i,A}}{N_A} \right) + N_{i,B}\left(\widetilde{\mu_i^0} + k_B T \log  \frac{N_{i,B}}{N_B} \right) \nonumber \\
&+ N_{s,A} \left(\widetilde{\mu_s^{0,*}} + k_B T \log \frac{N_{s,A}}{N_A} \right) + N_{s,B} \left(\widetilde{\mu_s^{0,*}} + k_B T \log  \frac{N_{s,B}}{N_B} \right) 
\label{eqa26} 
\end{align}  
and the final Gibbs free energy is 
\begin{align}
G^\text{final} &= \left(N_{i,A}-1\right)\left(\widetilde{\mu_i^0} + k_B T \log  \frac{N_{i,A}-1}{N_A-1} \right) + \left(N_{i,B}+1\right)\left(\widetilde{\mu_i^0} + k_B T \log  \frac{N_{i,B}+1}{N_B+1} \right) \nonumber \\
&+ N_{s,A} \left(\widetilde{\mu_s^{0,*}} + k_B T \log  \frac{N_{s,A}}{N_A-1} \right) + N_{s,B} \left(\widetilde{\mu_s^{0,*}} + k_B T \log \frac{N_{s,B}}{N_B+1} \right). 
\label{eqa27} 
\end{align} 

The difference of these two expressions leads to an exact formula, similar to Equation \ref{eqa13}, which we omit here.  Analogously to the approximation $N \gg \sigma$ made in the context of chemical reactions, here we assume that $N_A, \ N_B \gg 1$, which amounts to setting $N_A -1 \approx N_A$ and $N_B +1 \approx N_B$ in Equations \ref{eqa26} and \ref{eqa27}.  Using this approximation, we can express the change in Gibbs free energy as 
\begin{equation}
\Delta G = k_B T \log \frac{(N_{i,A}-1)^{(N_{i,A}-1)}}{N_{i,A}^{N_{i,A}}} \frac{(N_{i,B}+1)^{(N_{i,B}+1)}}{N_{i,B}^{N_{i,B}}}
\label{eqa28}
\end{equation}
We see that we, analogously to making the approximation $N_i \gg \nu_i$ above, if we here make the approximation $N_{i,A}, N_{i,B} \gg 1$, we can reduce Equation \ref{eqa28} to the textbook expression 
\begin{equation}
\Delta G = k_B T \log \frac{N_{i,B}}{N_{i,A}}.
\label{eqa29}
\end{equation}
The right hand side of Equation \ref{eqa29} will always be less than that of Equation \ref{eqa28}, which, although the difference is very slight, can lead to systematically biased calculations, as we show next.  

\section{Results}
Here we perform simple stochastic simulations to illustrate the effects of approximating $\Delta G$ for reactions and diffusion when copy numbers are considered small and discrete.  We use MEDYAN, a simulation platform useful for studying active networks at high resolution which is equipped with a RDME simulation engine of the type described above \cite{popov2016medyan}.  With this, we report on two different simulations, to test the discrepancy between the nearly exact and approximate forms of $\Delta G$ corresponding to reactions and to diffusion.  We compare the nearly exact Equations \ref{eqa16} and \ref{eqa28} for reactions and diffusion, respectively, with their approximate counterparts, Equations \ref{eqa21} and \ref{eqa29}.  We observe that only the nearly exact expressions result in sensible behavior, i.e. that the rate of change of Gibbs free energy on average is 0 $k_B T / s$ when the system is at equilibrium.

\begin{figure}[H]
	\centering
	\includegraphics[width= 15 cm]{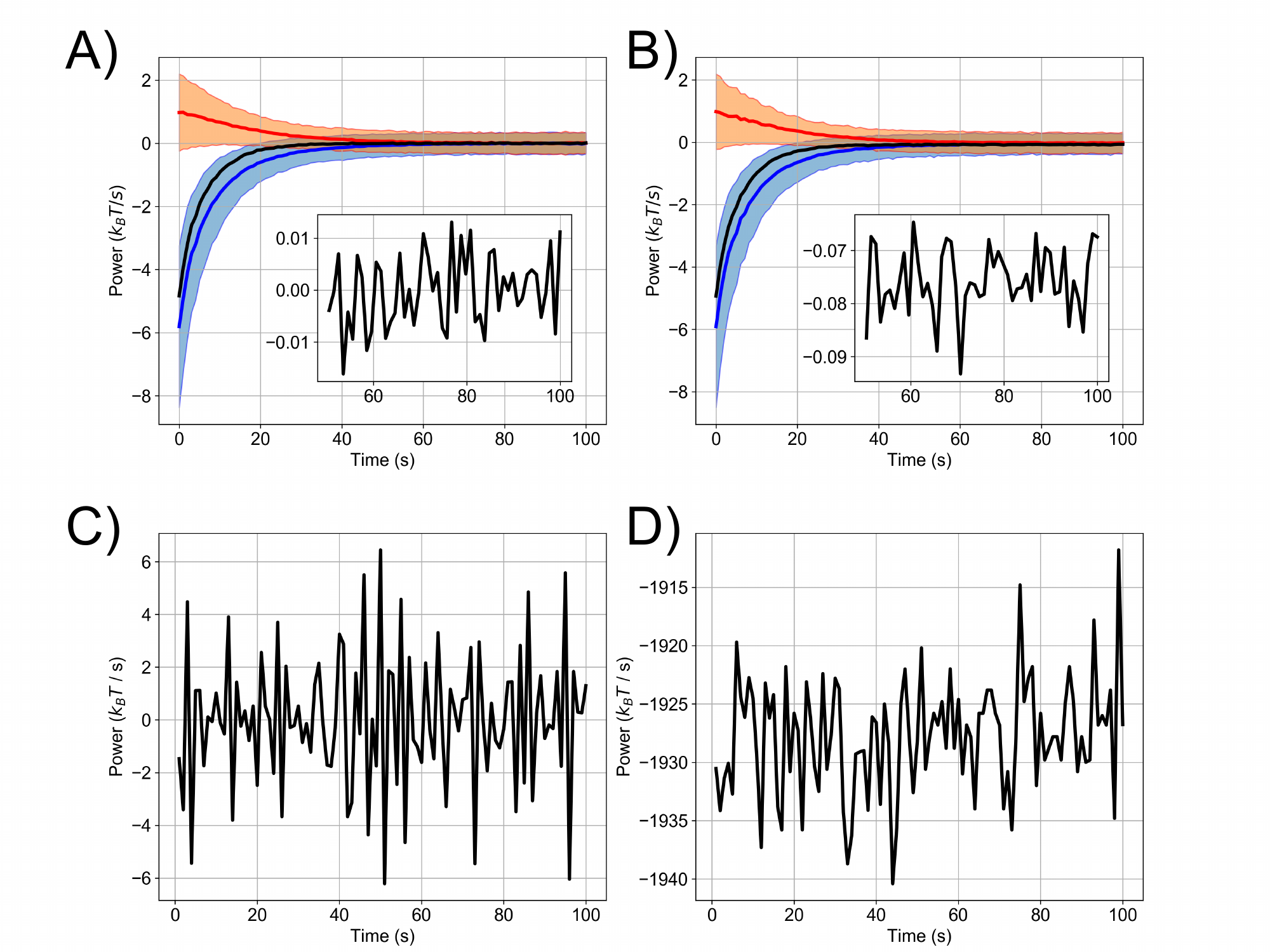}
	\caption{Numerical results illustrating the difference between nearly exact and approximate formulations of $\Delta G$ for reactions and diffusion.  A) Averages over repeated simulations of the chemical scheme described by Equation \ref{eqa30}, using the nearly exact Equation \ref{eqa16}.  The blue curve represents the trajectory of $\partial_t \Delta G$ resulting from the forward reaction, the red curve represents that from the reverse reaction, and the black curve represents their sum.  Shaded regions represent the standard deviation over the 3,000 repeated trials.  The inset displays a blow-up of the black curve once the system has gotten close to equilibrium.  B) The same as just described, but using the approximate Equation \ref{eqa21}.  C) A single trajectory of the diffusion of 1,000 molecules over a 1 $\mu m^3$ cubic grid of 8 compartments, beginning from a random initial spatial distribution.  Values of $\Delta G$ are calculated using Equation $\ref{eqa28}$.  D) The same as just described, however values of $\Delta G$ are calculated using Equation $\ref{eqa29}$.    }
	\label{abcdiff}
\end{figure}

To test the effect of approximation for $\Delta G$ of reactions, we consider the following simple reaction scheme:
\begin{equation}
A + B \leftrightharpoons C
\label{eqa30}
\end{equation}
where the rate constant to the right is $k_f = 0.05$ $\mu M^{-1} s^{-1}$, and the rate constant to the left is $k_r = 0.01$ $s^{-1}$, giving an equilibrium constant of $K_{eq} = 0.2$ $\mu M$.  We perform stochastic simulations with the Next Reaction Method (NRM) in MEDYAN \cite{gibson2000efficient}, using a single compartment of size $0.125$ $\mu m^3$ (i.e. in this example there is no diffusion).  To employ Equation \ref{eqa16} in simulation, we first compute $\Delta G^0 - \sigma k_B T \log {\Theta} - \sigma k_B T = 3.71 \ k_B T$ for the forward reaction and $-3.71 \ k_B T$ for the reverse reaction, and then when each reaction fires during simulation, the quantity $\widetilde{Q}_{(1)}$ is computed from the instantaneous values of the copy numbers to determine $\Delta G_{(1)}$.  A similar approach is taken to employ Equation \ref{eqa21}.  We begin with $N_A = 100$, $N_B = N_C = 50$, and repeat a simulation of 100 $s$ duration 3,000 times to obtain averages of the trajectory of rates $ \partial _t \Delta G (t)$ resulting from the forward and reverse reactions.  Figure \ref{abcdiff} displays the results of these simulations.  Note how, while the two trajectories bear close similarity, the equilibrium value of the total rate of $\Delta G$ centers around 0 $k_B T / s$ for the nearly exact formulation, yet erroneously centers around $\sim$ $-0.08$ $k_B T/s$ for the approximate version.  
	
To test the effect of approximating $\Delta G$ for diffusion in a compartment-based reaction-diffusion scheme, we next employed MEDYAN to simulate diffusion of 1,000 molecules with diffusion constant $20$ $\mu m^2 s^{-1}$ in a $2 \times 2 \times 2$ grid of compartments, each a cube with volume $0.125$ $\mu m^3$.  The initial distribution of molecules is uniformly random over the compartment grid, and thus the system begins at equilibrium and is then allowed to stochastically evolve for 100 $s$, i.e. the molecules hop randomly between adjacent compartments.  For each hopping reaction, the value of $\Delta G$ is determined using the nearly exact Equation \ref{eqa21} for one run, and in another run the approximate Equation \ref{eqa29} is used.  The difference in the trajectories of of $\partial_t \Delta G$ for these simulations is stark, as displayed in Figure \ref{abcdiff}.  While the trajectory centers around 0 $k_B T/s$ for the nearly exact formulation of $\Delta G$, it erroneously centers around $\sim$ $-1,930$ $k_B T/s$ for the approximate version.  Diffusion events are very frequent in this system, occurring around 240,000 times per second (this number can be calculated from the parameters of the system and is also observed during simulations).  Thus, since Equation \ref{eqa29} is always less than the nearly exact quantity, even by a small amount on the order 0.05 $k_B T$ for this system, this systematic bias is amplified by the frequency of diffusion events to produce significant differences from expected behavior, necessitating the use of a more exact formula for $\Delta G$. 
Lastly, we performed a simulation involving both reactions and diffusion across multiple compartments.  We found again that only when the nearly exact formulas were used did the rate of change of Gibbs free energy center on 0 $k_B T / s$ at equilibrium.  It is not additionally illuminating to show the data, so we do not display it here.    

\section{Discussion}
\begin{figure}[H]
	\centering
	\includegraphics[width= 15 cm]{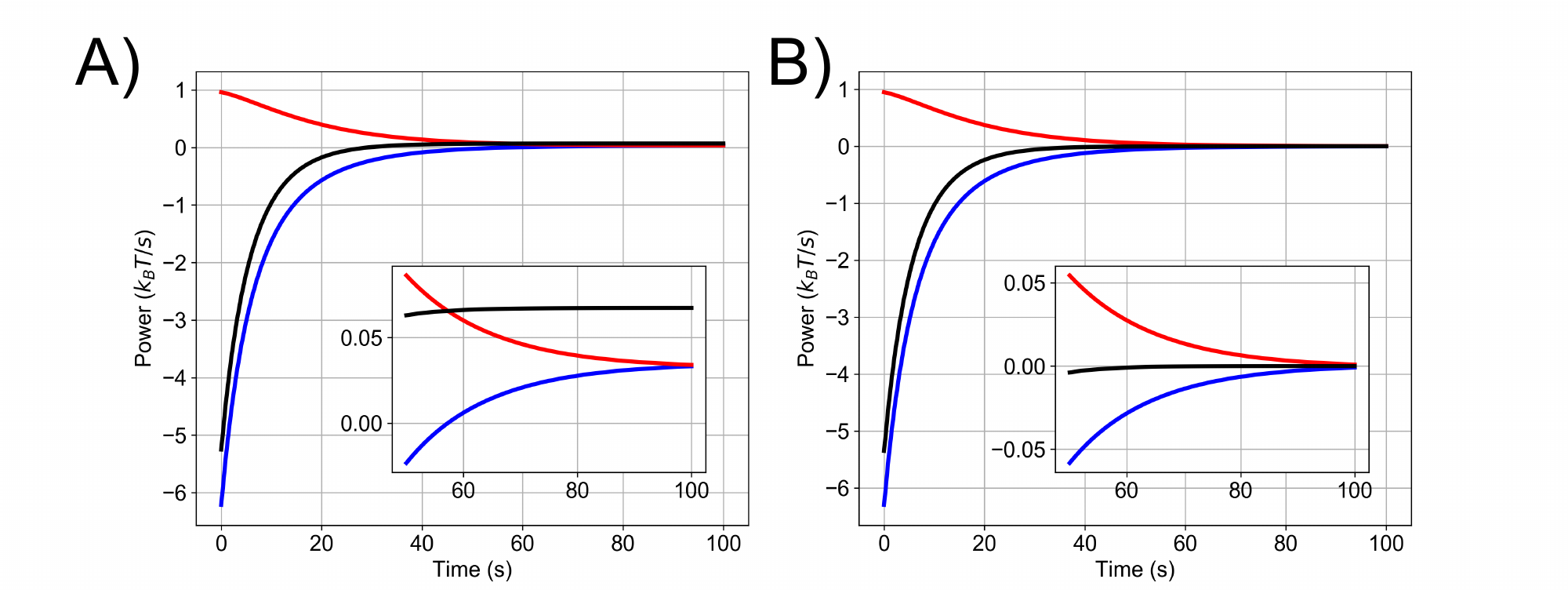}
	\caption{Analytical results illustrating that, when copy numbers are treated as continuous variables, the textbook formula for $\Delta G$ of reaction should be used.  A) Calculation of the rates $\partial_t \Delta G(t)$ using Equation \ref{eqa16} resulting from the forward (blue curve) and reverse (red curve) reactions, as well as their sum (black curve), as described in the main text.  The inset shows a blow up of these curves as the system nears equilibrium.  B) The same as just described, but using Equation \ref{eqa21}.  }
	\label{mfcomb}
\end{figure}

We have argued that when the copy numbers of the reactants and products are treated as small, discrete quantities, then certain approximations leading to the textbook formulas for $\Delta G$ of reaction and diffusion, Equations \ref{eqfirst} and \ref{eqsecond}, break down and lead to biased results.  We emphasize that this is true only when the copy numbers and reaction occurrences are treated as discrete; in fact when they are treated as continuous, one should use the textbook formulas.  This can be shown by considering a continuous version of the chemical system described by Equation \ref{eqa30}.  The time evolution of the concentrations of the chemical species is obtained by solving a system of ordinary differential equations that employ mass-action kinetics, and from this solution, the rates $\partial_t \Delta G(t)$ resulting from the forward and reverse reactions are computed using the instantaneous values of the species' copy numbers which enter into Equations \ref{eqa16} and \ref{eqa21}.  Figure \ref{mfcomb} displays the results of these calculations.  Here, the total rate of $\partial_t \Delta G(t)$ only approaches 0 $k_B T \ / s$, as it must at equilibrium, when Equation \ref{eqa21} is used.   

When the copy numbers of the chemical species are treated as continuous, there is no notion of a single occurrence of a reaction; instead the evolution of the system is parameterized by the continuous variable $\xi $ which quantifies the extent of advancement of the reaction \cite{borge2014reviewing}.  In this framework, which is adopted in classical thermodynamics, the copy number of any species never jumps instantaneously from $N_i$ to $N_i + \nu_i$, and thus the premise of the derivation presented above leading to Equation \ref{eqa13} falls apart.  This explains why the seemingly more accurate Equation \ref{eqa16} is wrong when applied to chemical dynamics that are described using continuous variables to represent the copy number of species.  When the chemical dynamics are modeled this way, the textbook expression Equation \ref{eqa21} for the change in Gibbs free energy is valid.  

However, if the copy numbers of chemical species are not large compared to the stoichiometric coefficients, a better description of the chemical dynamics is found by treating the copy numbers as discrete variables participating in stochastically timed chemical reactions.  This is the philosophy adopted by several recent models of intracellular environments, where copy numbers of molecules of interest are sometimes quite small.  In these cases, adoption of the more exact expression for $\Delta G$ can not only improve precision, but can ensure correct behavior.   Particularly when diffusion down concentration gradients is expected to be a strong source of entropy production (e.g. diffusion of monomeric actin along the lengths of filopodia, see \cite{erban2014multiscale, dobramysl2016steric}), then using the nearly exact formula presented here can ensure that the resulting measurement of dissipation are not strongly biased.

\section*{Appendix A: Discrete Variables and the Relation to the Gibbs-Duhem Equation}

The Gibbs-Duhem Equation from classical thermodynamics states 
\begin{equation}
\sum_{i \in M} N_i d \mu_i = -S dT + V dp = 0
\label{eqa34}
\end{equation}
where $M = R \cup P \cup s$ represents the set of all chemical species in the system including the solvent, and where the last equality holds at constant temperature and pressure \cite{callen1998thermodynamics}.  Applying the product rule\footnote{As a reminder, the product rule can be derived as $d (fg) = (f + df)(g + dg) - fg = fdg + g df + df dg = f dg + g df$ since the term $df dg$ is assumed to be negligible.} to the expression for the Gibbs free energy, $G = \sum_{i \in M} N_i \mu_i$, gives

\begin{equation}
d G = \sum_{i \in M} \mu_i d N_i  + \sum_{i \in M} N_i d \mu_i = \sum_{i \in M} \mu_i d N_i .
\label{eqa35}
\end{equation}
Evaluating $\sum_{i \in M} \mu_i d N_i $ using stoichiometric coefficients in place of $d N_i$ can be shown to lead to the expression $\Delta G_{(3)}$.  

If the copy numbers are small, we are better served using the discrete difference operator $\Delta$ and not the differential difference operator $d$.  The product rule for the discrete difference operator applied to $G $ gives
\begin{equation}
\Delta G = \sum_{i \in M} \mu_i \Delta N_i + \sum_{i \in M} N_i \Delta \mu_i + \sum_{i \in M} \Delta \mu_i \Delta N_i.  
\label{eqa36}
\end{equation}
Note that here we do not neglect the cross-term as we do in the differential product rule.  We can evaluate this expression term by term:
\begin{equation}
\sum_{i \in M} \mu_i \Delta N_i = \Delta G_{(3)}
\label{eqa37}
\end{equation}
\begin{equation}
\sum_{i \in M} N_i \Delta \mu_i =  k_B T \log \prod_{i \in R} \left( \frac{N_i - \nu_i}{N_i} \right) ^{N_i}  \prod_{j \in P}\left( \frac{N_i + \upsilon_j}{N_j} \right) ^{N_j}  \left( \frac{N}{N+\sigma}\right) ^N 
\label{eqa38}
\end{equation}

\begin{equation}
\sum_{i \in M} \Delta  N_i \Delta \mu_i =  k_B T \log  \prod_{i \in R} \left( \frac{N_i - \nu_i}{N_i} \right) ^{-\nu_i}  \prod_{j \in P}\left( \frac{N_i + \upsilon_j}{N_j} \right) ^{\upsilon_j}  \left( \frac{N}{N+\sigma}\right) ^\sigma  
\label{eqa39}
\end{equation}
Comparing these to the expressions for the accuracy of the various approximations given in Appendix B, we have 
\begin{equation}
\sum_{i \in M}   N_i \Delta \mu_i + \sum_{i \in M} \Delta  N_i \Delta \mu_i = \Delta G_{(0)} - \Delta G_{(3)}
\label{eqa40}
\end{equation}
and thus, combining all the terms in Equation \ref{eqa36}, we recover our exact expression $\Delta G_{(0)}$.  To summarize, for small, discrete copy numbers, we obtain corrections to the expression for the change in Gibbs free energy accompanying chemical reactions that are not captured by the constraints imposed by the Gibbs-Duhem Equation, which assumes that the copy numbers are continuous quantities.

\section*{Appendix B: Accuracy of the Approximations} 
We compute the accuracy of the approximations $\Delta G_{(1)}, \Delta G_{(2)}$, and $\Delta G_{(3)}$ by taking the difference of these quantities with $\Delta G_{(0)}$.  We have 
\begin{equation}
(\Delta G_{(1)} - \Delta G_{(0)})/k_BT =  \log \left(\frac{N+\sigma}{N}\right)^{N+\sigma} - \sigma 
\label{eqa31}
\end{equation}
\begin{equation}
(\Delta G_{(2)} - \Delta G_{(0)})/k_BT =  \log \left(\frac{N+\sigma}{N}\right)^{N+\sigma} + \log \prod_{i \in R} \left(\frac{N_i}{N_i - \nu_i}\right)^{N_i} \prod_{j \in P} \left(\frac{N_j}{N_j + \upsilon_j}\right)^{N_j}   
\label{eqa32}
\end{equation}
\begin{equation}
(\Delta G_{(3)} - \Delta G_{(0)})/k_BT =  \log \left(\frac{N+\sigma}{N}\right)^{N+\sigma} + \log \prod_{i \in R} \left(\frac{N_i}{N_i - \nu_i}\right)^{N_i-\nu_i} \prod_{j \in P} \left(\frac{N_j}{N_j + \upsilon_j}\right)^{N_j + \upsilon_j}    
\label{eqa33}
\end{equation}
We see that the accuracy of $\Delta G_{(1)}$ depends only on $N$ for a given reaction, whereas the remaining approximations depend also on the values of $N_i$ and $N_j$.  These observations are consistent with the fact that, in order to arrive at the expression for $\Delta G_{(1)}$, we leveraged the size of $N$ compared to $\sigma$, and to arrive at the expressions for $\Delta G_{(2)}$ and $\Delta G_{(3)}$ we further successively leveraged the  sizes of $N_i, N_j$ compared to $\nu_i, \nu_j$.  

\section*{Appendix C: $\Delta G$ of Solvent Fluctuations}
Here we first calculate an approximate expression for the change in Gibbs free energy of the system accompanying a fluctuation of $n$ solvent molecules from some compartment A to compartment B.  We have 
\begin{eqnarray}
G^\text{initial}  &=& \sum_{i \in S} N_{i,A}\left(\widetilde{\mu_{i}^0} + k_B T \log{\frac{N_{i,A}}{N_A}}\right) + N_{s,A}\left(\widetilde{\mu_s^{0,*}} + k_B T \log{\frac{N_{s,A}}{N_A}}\right) \nonumber \\
 &&+ \sum_{i \in S} N_{i,B}\left(\widetilde{\mu_{i}^0} + k_B T \log{\frac{N_{i,B}}{N_B}}\right) + N_{s,B}\left(\widetilde{\mu_s^{0,*}} + k_B T \log{\frac{N_{s,B}}{N_B}}\right),
 \label{eqx1}
\end{eqnarray}
where $S$ is the set of solute species.  After the transfer of $n$ solvent molecules from $A$ to $B$ we have
\begin{eqnarray}
G^\text{final}  &=& \sum_{i \in S} N_{i,A}\left(\widetilde{\mu_{i}^0} + k_B T \log{\frac{N_{i,A}}{N_A-n}}\right) + N_{s,A}\left(\widetilde{\mu_s^{0,*}} + k_B T \log{\frac{N_{s,A}-n}{N_A-n}}\right) \nonumber \\
&&+ \sum_{i \in S} N_{i,B}\left(\widetilde{\mu_{i}^0} + k_B T \log{\frac{N_{i,B}}{N_B+n}}\right) + N_{s,B}\left(\widetilde{\mu_s^{0,*}} + k_B T \log{\frac{N_{s,B}+n}{N_B+n}}\right)
\label{eqx2}.
\end{eqnarray}
Taking the difference, and simplifying, we arrive at the exact expression
\begin{equation}
\Delta G = k_B T \log{\frac{N_A^{N_A}}{(N_A - n)^{N_A - n}}
	\frac{(N_{s,A }- n)^{N_{s,A }- n}}{N_{s,A}^{N_{s,A}}}
	\frac{N_B^{N_B}}{(N_A + n)^{N_A + n}}
	\frac{(N_{s,B}+ n)^{N_{s,B}+ n}}{N_{s,B}^{N_{s,B}}}}.
\label{eqx3}
\end{equation}
To understand the magnitude of this expression for typical values of $n$ and $N_{s,A}$ compared to $N_A$, we first make the assumption that the two compartments initially have the same number of solvent molecules, i.e. that  $N_{s,A} = N_{s,B} \equiv N_s$.  Next we assume that the solvent molecules dominate the proportion of total molecules, allowing us to write $N_A \approx N_B \equiv N$. These approximations will hold in the limit that the number of solute molecules is much less than the number of solvent molecules for each compartment.  We next introduce the small parameters 
\begin{equation}
\epsilon_A = \frac{\sum_{i \in S}N_{i,A}}{N},
\label{eqx4a}
\end{equation}
and
\begin{equation}
\epsilon_B = \frac{\sum_{i \in S}N_{i,B}}{N},
\label{eqx4}
\end{equation}
which captures the dilution of two compartments, and 
\begin{equation}
\xi = \frac{n}{N},
\label{eqx5}
\end{equation}
which represents the comparative size of the fluctuation.  For $N = 10^9$, then typically $\xi \sim 10^{-4.5}$ and $\epsilon \sim 10^{-6}$.  
Substituting in these parameters, we have
\begin{eqnarray}
\Delta G &=& k_B T \log \frac{N^{2N}}{(N(1-\xi))^{N(1-\xi)}(N(1+\xi))^{N(1+\xi)}} \times \nonumber \\
&& \frac{(N(1-\epsilon_A - \xi))^{N(1-\epsilon_A - \xi)}(N(1-\epsilon_B + \xi))^{N(1-\epsilon_B + \xi)}}{(N(1-\epsilon_A))^{N(1-\epsilon_A)}(N(1-\epsilon_B))^{N(1-\epsilon_B)}}
\label{eqx6}
\end{eqnarray}
We first expand this expression to linear order in $\epsilon_A$ and $\epsilon_B$, and then to second order in $\xi$.  The result is
\begin{eqnarray}
\Delta G &\approx&   (\epsilon_A - \epsilon_B) N \xi k_B T  + \frac{1}{2} (\epsilon_A + \epsilon_B) N \xi^2 k_B T \nonumber \\
&=& (\epsilon_A - \epsilon_B) n k_B T + \frac{1}{2} (\epsilon_A + \epsilon_B) \frac{n^2}{N} k_B T
\label{eqx7}
\end{eqnarray}

The observed numerical agreement between Equations \ref{eqx3} and Equations \ref{eqx7} is close for realistic values of the parameters: for $N_A = 10^9$, $N_B = 5\times 10^8$, $\epsilon_A = 10^{-6}$, $\epsilon_B = 3 \times 10^{-6}$, and $n = N_A^{1/2}$, the prediction of Equation \ref{eqx3} is $-0.0632401 \ k_B T$, and the prediction of Equation \ref{eqx7} is $-0.0632436 \ k_B T$.  The first term in Equation \ref{eqx7} will dominate if the solute dilutions in the two compartments are very different from each other.  In this case, we may compare the size of this change in Gibbs free energy to that accompanying the diffusion of a solute from compartment B to compartment A.  This latter change in Gibbs free energy will be approximately $k_B T \log \frac{\epsilon_A}{\epsilon_B}$.  If we now set $\epsilon_A = a \epsilon_B$, where $a$ is of order 1 (typically it will fall in the range $[\frac{1}{10}, 10]$), then $\Delta G$  for the solvent fluctuation will be $k_B T (a-1) \epsilon_B n$ and $\Delta G$ for the solute diffusion will be $k_B T \log{a}$.  The product $\epsilon_B n$ will typically on the order of $\sim 10^{-1.5}$, so one can see that for usual values of the parameters the change in Gibbs free energy from a solute diffusion event will be greater in magnitude than that from a solvent fluctuation.  If the dilutions are very similar, then $\epsilon_A - \epsilon_B \approx 0$, and the second term in Equation \ref{eqx7} dominates.  This term is on the order of $(\epsilon_A + \epsilon_B) k_B T$ since $\frac{n^2}{N} \sim 1$.  These changes in Gibbs free energy will typically be much smaller than those accompanying a chemical reaction or inter-compartment diffusion of the solute, and thus we may neglect the activity of the solvent in tracking the Gibbs free energy of the system.

\section*{Funding}
This work was supported by the following grants from the National Science Foundation: NSF 1632976, NSF DMR-1506969, and NSF CHE-1800418.
\section*{Acknowledgments}
The authors gratefully acknowledge A. Chandresekaran, H. Ni, and Q. Ni for their improvements of the manuscript and helpful discussions.

\bibliographystyle{unsrt}

\end{document}